# Ultra-Broadband Visible and Infrared Light Generation Driven by Far Infrared Light in the Broad Region from 8 µm to 240 µm

Nils W. Rosemann[1,2,*], Robin C. Döring[2,3], Eike Dornsiepen[4,5], Jürgen Belz[3], Johannes Haust[3], Felix Bernhardt[6], Ferdinand Ziese[6], Claudio Attaccalite[7,8], Stephan Winnerl[9], Harald Schneider[9], Manfred Helm[9], Stefanie Dehnen[4], Kerstin Volz[3], Simone Sanna[6], and Sangam Chatterjee[2]

[1]*Light Technology Institute, Karlsruhe Institute of Technology (KIT), D-76131 Karlsruhe, Germany*

[2]*Institute of Experimental Physics I and Center for Materials Research (ZfM/LaMa), Justus-Liebig-University Giessen, Heinrich-Buff-Ring 16, 35392 Gießen, Germany*

[3]*Fachbereich Physik und Wissenschaftliches Zentrum für Materialwissenschaften, Philipps-Universität Marburg, Hans-Meerwein-Straße, 35043 Marburg, Germany*

[4]*Fachbereich Chemie und Wissenschaftliches Zentrum für Materialwissenschaften, Philipps-Universität Marburg, Hans-Meerwein-Straße, 35043 Marburg, Germany*

[5]*Faculty of Chemistry, University of Duisburg-Essen, Universitätsstraße 5-7, 45117 Essen, Germany*

[6]*Institute of Theoretical Physics and Center for Materials Research (ZfM/LaMa), Justus-Liebig-University Giessen, Heinrich-Buff-Ring 16, 35392 Gießen, Germany*

[7]*CNRS/Aix-Marseille Université, Centre Interdisciplinaire de Nanoscience de Marseille UMR 7325 Campus de Luminy, 13288 Marseille Cedex 9, France*

[8]*CNR-ISM, Division of Ultrafast Processes in Materials (FLASHit), Area della Ricerca di Roma 1, Via Salaria Km 29.3, I-00016 Monterotondo Scalo, Italy*

[9]*Institute of Ion Beam Physics and Materials Research, Helmholtz-Zentrum Dresden-Rossendorf, Bautzner Landstraße 400, 01314 Dresden, Germany*

The availability of tailored, spectrally broadband "white-light" is instrumental for various spectroscopic or sensing applications, and hence remains a very active field-of-research. In particular low-étendue, high brightness, brilliant continuum sources are challenging. Here we show that compounds comprising adamantane-type cluster molecules allow the generation of a spectrally broad continuum for an exceptionally broad range of pump-wavelengths spanning all the way from the far infrared towards the visible. By using the tunable emission of a free-electron laser, we are able to achieve a broadband emission spanning from 9 µm to 500 nm for pump-wavelengths in the range from 8 µm to 240 µm. Thus, outperforming the presumably most established supercontinuum sources platform based on photonic crystal fibers, where the specific fiber design commonly confines its operation to only one selected narrow pump wavelength. The conversion efficiency of the observed cluster-molecule based continuum generation also depends on the driving wavelength. Detailed theoretical calculations are used to identify infrared-active core modes in the investigated range of pump-wavelengths. In this way, we clearly identify vibrational motion – and hence Joule heating – of the cluster core as energy dissipation mechanism competing with continuum emission. Furthermore, we underline the significant role that the clusters electronic system plays in the observed continuum emission.

# I. INTRODUCTION

The field of nonlinear optics became of scientific interest already shortly after the realization of the laser in 1960 and has rapidly been advancing ever since. Nowadays, nonlinear optical effects are widely employed [1]. Prominent examples include low-order nonlinear processes like second-harmonic generation (SHG) for green, or sum-frequency generation for orange/yellow laser pointers. More extreme nonlinear effects such as high-harmonic generation provide attosecond pulses in the deep ultra-violet or soft x-ray regime [2–4]. Supercontinuum generation is used for chemical sensing, coherent anti-Stokes Raman spectroscopy, optical coherence tomography amongst others [5–8].

In general, the term supercontinuum is used widely and, unfortunately, rather loosely as it lacks a strict unambiguous definition. It very often refers to a spectrum spanning at least one octave of frequencies without any major spectral drops. The specific wavelength region, however, is not defined. Thus, a plethora of phenomena fall under this loose definition and an accordingly large number of publications report on supercontinua ranging from the visible spectral range to the infrared and even beyond [9–11]. Similar to the variety of wavelength regions, a vast assortment of techniques is used to generate supercontinua. A rough categorization in terms of nonlinear media ranges from laser-generated plasmas to condensed matter systems. The latter can be further divided into a multitude of sub-systems such as bulk solids or liquids, as well as specifically designed nanostructures or fibers, either photonic crystal or tapered [8,12,13].

Many extreme nonlinear optical effects require high peak intensities delivered by ultrafast lasers. For example, high-harmonic generation in plasmas is only invoked by electric field strength capable of transient ionization of gases [14,15]. Such capable laser sources render this method quite expensive. Alternative nonlinear processes in bulk materials such as beta-barium borate circumvent ionization steps, but still require rather large electric field strengths [12,16–18]: here, the field strength are mandated by the rather low nonlinear coefficients, limited by the intrinsic refractive index of the nonlinear optical material [19–22]. This limit can be surpassed by designing tailored photonic structures that exhibit extreme nonlinear coefficients, e.g., waveguides or photonic crystal fibers [23–25]. Such structures provide two main benefits: they confine the transported light to a small mode area which is inversely proportional to the maximum accessible phase shift, i.e., spectral broadening of the pump pulse [19]. Additionally, these structures result from specially designed profiles of refractive indices by the combination of different media, including vacuum as one medium. This allows for a targeted design of the dispersion, in particular, the zero-dispersion wavelength and the so-called anomalous dispersion regime. While the zero-dispersion wavelength is important for fiber-based communication, the anomalous dispersion regime is often chosen for supercontinuum generation [8]. Due to the monotonic dispersion of most natural materials, the anomalous dispersion is often found only in the vicinity of an absorption resonance of the medium [26]. The anomalous dispersion can, thus, only be tailored within very limited material specific and geometry dependent wavelength limits. Additionally, the material's absorption limits the accessible wavelength range of the respective nonlinear medium. Beyond the linear absorption of the pump, this includes two-photon and higher-order photon absorption as well as reabsorption of the generated supercontinuum.

# II. CLUSTER MOLECULE-BASED CONTINUUM GENERATION

An intriguing yet comparatively unexplored mechanism of supercontinuum generation employs amorphous nonlinear media comprised of adamantane-type organotetrel-chalcogenide cluster molecules [27]. Here, a broad white-light supercontinuum covering at least the visible range was reported for pumping wavelength from 700 nm to 1100 nm, while crystalline material shows discrete harmonics rather than a broadband continuum [28]. The tentatively proposed mechanism in such amorphous systems involves the emission caused by accelerated charges, presumably in the electronic ground-state potential of the cluster molecules. In contrast to supercontinuum generation in fibers, this mechanism does not require the anomalous dispersion region. Simultaneously, it avoids the extreme field strength required for supercontinuum generation in gaseous media. The presumed usage of the electronic ground-state only, i.e., driving the system at energies below the optical absorption edge, avoids unwanted absorption that is expected to act as competing loss channel. This indicates a virtually unlimited range of driving frequencies towards low energies for the observed white-light generation. These could then potentially include electronic, radio-frequency or microwave sources.

Additionally, the assigned supercontinuum generation mechanism comprises no lower energy boundary to the emission as indicated within time-dependent density functional theory. Moreover, calculation of the nonlinear susceptibility coefficients $\chi^{(2)}$ and $\chi^{(3)}$ performed in this work reveal the onset of strong optical nonlinearities at photon energies well below the optical band gap (see Supplemental Material, Figure S8 and Figure S9).

This work aims to contribute towards understanding the mechanism and establish a relationship between supercontinuum generation and the atomistic structure of the functional material, as well as indicate potential limitations in the low-energy driving range. To this end, we investigate amorphous powder of organotin sulfide cluster [(PhSn)$_4$S$_6$] (Ph = phenyl, C$_6$H$_5$) with respect to their nonlinear response driven by mid- to far-infrared light. We primarily focus on the visible emission; however, we also observe the as of yet unexplored emission in the near infrared region.

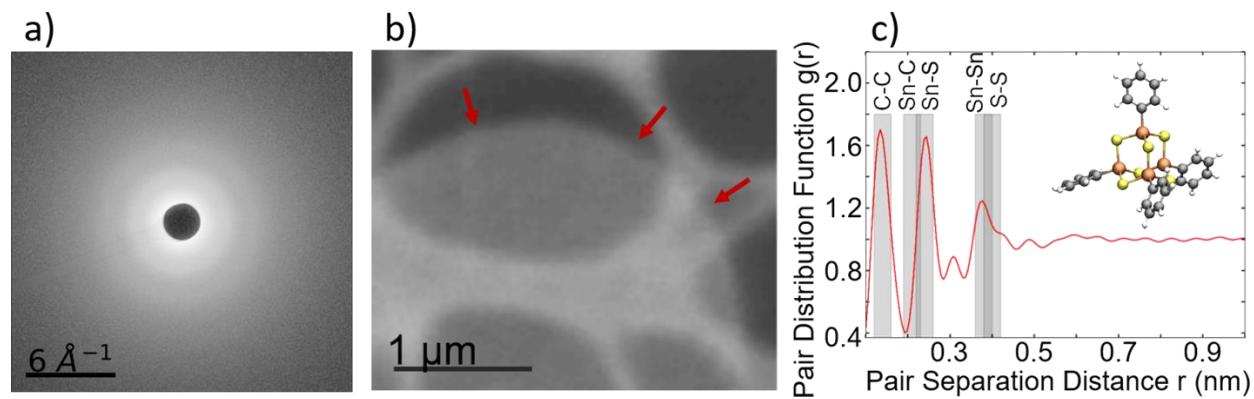

FIG. 1. a) Selected area diffraction pattern of a characteristic [(PhSn)$_4$S$_6$] sample showing the amorphousness of the sample. b) Virtual dark field from scanning precession electron diffraction data. The red arrows point towards the edge of one [(PhSn)$_4$S$_6$] powder. c) Pair distribution function with the positions of the theoretical atomic distances marked in grey. The molecular structure of [(PhSn)$_4$S$_6$] is shown as inset in c).

## A. Structural analysis

First, we quantify the amorphous structure of the material under investigation, performing a detailed structural analysis using sophisticated electron microscopic techniques. A powder of [(PhSn)$_4$S$_6$] is prepared according to the literature [29]. Powder particles are embedded in araldite epoxy and cured for subsequent ultramicrotomy cutting achieving slices as thin as 50 nanometers. These slices are transferred to cleaned, lacey-carbon coated electron microscopy grids. The selected area diffraction pattern of [(PhSn)$_4$S$_6$], exemplary shown in FIG. 1a), clearly underlines the amorphousness of the diffracting region. No diffraction spots stemming from crystalline areas have been found for any region of the specimen, which contains a multitude of powder particles. These amorphous particles can be detected by reconstructing a virtual dark field image from scanning precession electron diffraction data (FIG. 1b). Due to the larger scattering cross section of tin (Sn) and sulphur (S), the brighter [(PhSn)$_4$S$_6$] can be seen on the background of the carbon-based epoxy and the (web-like) lacey carbon TEM grid. Again, the homogeneous nature of the contrast in the image underlines the amorphousness of the material.

It should be noted that our compound consists of intact [(PhSn)$_4$S$_6$] molecular building blocks (shown in the inset of FIG. 1c) arranged in a random way in an amorphous solid. This is also reflected in the pair distribution function (FIG. 1c), derived from the diffraction pattern (FIG. 1a). Here we find peaks at the expected positions of the pair separation distance for characteristic atoms in the material.

## B. Visible emission

Having unambiguously established the amorphous structure, we now perform measurements of the nonlinear response of [(PhSn)$_4$S$_6$] using driving laser pulses in the mid-infrared to far-infrared spectral region. The experimental setup is sketched in FIG. 2, details are available in Supplemental Material.

Visible emission is observable with the bare eye for all evaluated pump-energies. It gives a warm white color

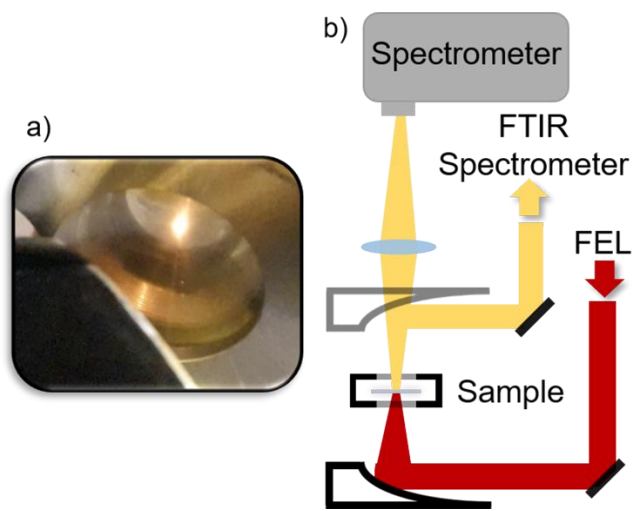

FIG. 2. a) Photograph of the sample in the vacuum chamber under excitation with pump wavelength of 20 µm (500 cm$^{-1}$). b) Sketch of the experimental setup. The light from the free-electron laser (FEL) is focused onto the sample inside the vacuum chamber. The emitted light is either collimated and relayed into an FTIR spectrometer or imaged onto a compact spectrometer. The collimating parabolic mirror is interchangeable and only inserted for detection with the FTIR spectrometer.

impression as shown in FIG. 2a, consistent with NIR excitation in previous reports [29]. The emission map for the whole available pump-energy range is given in FIG. 3a. An overview of the single excitation energies is given in the Supplemental Material. The emission maximum is found around 700 nm for the largest available pump photon energy. The emission undergoes a spectral shift to lower energies for decreasing pump photon energies. The emission maximum is found below 800 nm with an additional peak arising above 900 nm, i.e., at the edge of the experimentally accessible detection window above a pump energy of ~300 cm$^{-1}$ (34 µm). The observed shift in emission maximum is, however, not continuous but undergoes a discrete jump at ~300 cm$^{-1}$. Selected spectra are

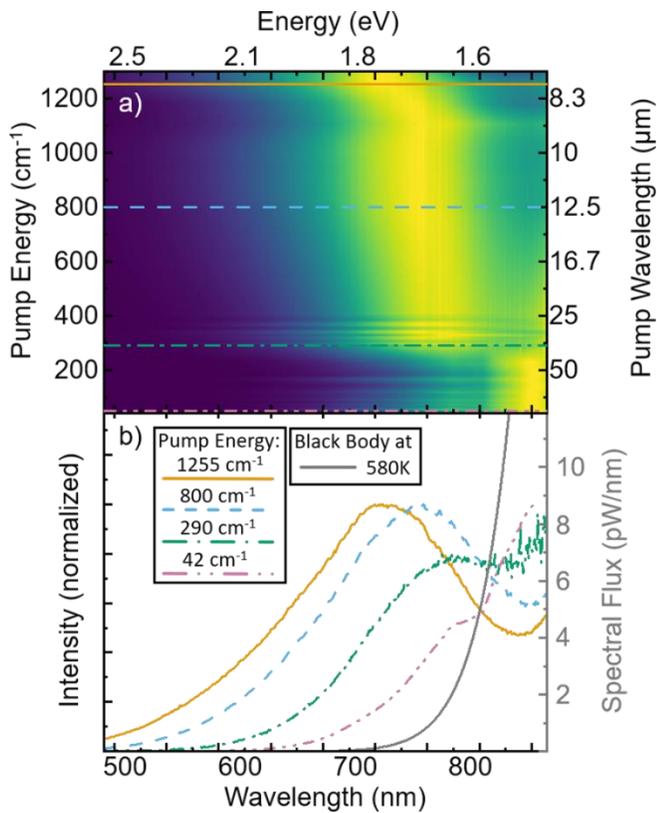

FIG. 3. Pump energy dependent visible emission of [(PhSn$_4$)S$_6$]. a) Linear false color map (dark to bright) of the VIS emission intensity for multiple excitation energies. b) Emission spectra for selected excitation energies (lines in a). For comparison, all spectra are normalized to their respective maximum in the detection window. The grey line indicates thermal emission from a black-body emitter at 580K.

The overall shape of the spectrum is governed by the detection system response and absorption caused by air. The latter is due to the fact that the beam path as well as IR detector are exposed to ambient conditions. The supercontinuum presented here appears to have very little structure by its own. This is caused by the underlying unique and highly nonlinear process that enables short yet effective interaction length of the pump pulse and the nonlinear medium, thus preventing soliton formation. Overall, the observation of visible white-light emission for pump-energies in the NIR and FIR region emphasizes the unparalleled nature of the underlying supercontinuum generation process. It also supports the previously assigned Bremsstrahlung-like nature, i.e., the emission of radiation by acceleration or deceleration of electrons. This is in stark contrast to spectral broadening due to, e.g., soliton formation. The latter strongly depends on the pump energy and typically features a spectral broadening depending on material parameters, interaction length, and incident field strength. Typically, this invokes broadenings spanning one to two octaves around the pump frequency. The observed emission up to 600 nm (500 THz) for pumping at 200 cm$^{-1}$ (6 THz) spanning more than six octaves below the pump frequency exceeds these apparent intrinsic limitations.

shown in FIG. 3b to visualize this non-monotonic evolution. The emission shows an additional shoulder below 800 nm for the pump energies where the emission maximum is found towards the end of the detection window. This indicates that the change in the emission maximum is more related to a change in ratio of different emission maxima, than to a continuous shift. Note that all visible emission spectra exceed the expected thermal emission from a perfect black-body emitter at 580 K, i.e., beyond the destruction threshold of the compound [29].

### C. Infrared emission

Next, we investigate the IR part of the emission. The spectral distribution of the emitted radiation for a driving photon energy of 287 cm$^{-1}$ (35 µm) measured by FTIR spectrometer is shown in FIG. 4. Unfortunately, the availability of beam time as well as comparably long acquisition times of the IR spectra limit the number of recorded IR spectra and no spectral map similar to Fig. 3 can be provided here. Nevertheless, all recorded IR spectra show a broad emission that extends all over the accessible detection window, from the visible to ~9 µm (~1100 cm$^{-1}$).

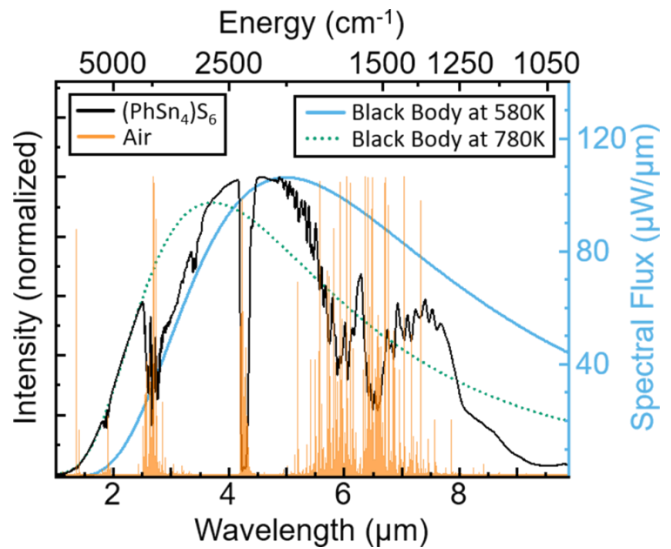

FIG. 4. Infrared emission of [(PhSn$_4$)S$_6$]. Emission (black line) observed for an excitation energy of 286.9 cm$^{-1}$ (34.86 µm) for 6.9 nJ/pulse and peak excitation density of 17.6 kW/cm$^2$. The spectral shape is not corrected for detection system response or absorption caused by moisture in the air. For reference the absorption of air (orange) is given and the high energy flank is fitted by a black-body emitter at 780 K. Additionally the emission spectrum of a black-body emitter at 580 K is given (blue line) with the spectral flux corresponding to a circular emitter of 0.2 mm radius.

The assignment to a Bremsstrahlung-like process gets further corroborated when considering the infrared part of the emission spectrum.

### D. Excitation dependent oscillations

In addition to the overall peak shift, the emission intensity apparently strongly depends on the driving photon energy, as shown by the distinct intensity fluctuations in Fig 2a, particularly below 400 cm$^{-1}$. To highlight this observation, the integrated emitted intensity is plotted versus pump energy in FIG. 5a. The vertical error bars are based on averaging multiple measurements at various sample positions. This ensures that the fluctuations are neither caused by experimental parameters such as pump intensity fluctuations, nor by sample inhomogeneities. The horizontal error bars represent the full-width at half-maximum of the FEL at the respective wavelength. The intensity fluctuations are significantly larger than the experimental uncertainty, in particular for pump-energies <250 cm$^{-1}$. Here, the pump-energy is in the range of typical molecular vibrational modes. Dissipation of pump energy into such vibrations could act as competing channel to the continuum generation and inhibit the latter. Vibrational spectra calculated by first-principles calculations corroborate this hypothesis. Details of the density-functional theory approach are given in the Supplemental Material. The calculations are successfully benchmarked against experimental data for energies below 2000 cm$^{-1}$ (See Figure S1). The calculated infrared (vibrational) absorption modes including a constant Gaussian broadening of the individual intensities are shown as grey-shaded areas in FIG. 5a. Strikingly, virtually all calculated absorption resonances coincide with minima in the emitted intensity. This clearly supports our hypothesis that the presence of additional loss channel competes or even inhibits the nonlinear optical response. This can be explained as follows: whenever the incident pump-beam is in resonance with a molecular vibration, a (potentially significant) fraction of the pump beam is absorbed by the molecules and, thus, cannot contribute to the optical response. For the other pump-energies non-resonant to vibrational transitions, the incident pump can then accelerate the electron system. The extreme detuning of the incident pump photons from the HOMO-LUMO transition renders multi-photon processes extremely unlikely. However, delocalized electrons in condensed matter are well known to couple to driving far-infrared fields in the ground state, [30] a process commonly leading to heating of the electronic system once equilibrium is reached amongst the charge carriers. [31] Tentatively, the electronic system appears to "cool" by Bremsstrahlung-like emission from the electronic system rather than by transferring the energy into vibrations of the molecular backbone. This mechanism is somewhat similar to reported ion acceleration accompanied by white-light emission without significant heat transfer into the crystal lattice [32].

### III. DISCUSSION

The numerical results reveal the microscopic nature of the motion associated with the infrared absorbing resonances. All strongly absorbing features that coincide with weaker visible emission are associated with motion of atoms in the cluster core (See animation in the Supplemental Material). This means that at such resonant pump-energies a significant amount of energy is transferred into the molecular back-bone itself, similar to Joule (resistive) heating. For temperatures above 580K, this leads to transformation of the compound, as observed by thermogravimetric analysis [29]. Even for a perfect black-body emitter, such a comparatively low quasi-equilibrium temperature is not sufficient for thermal emission to match the observed spectral emission distribution (cf. FIG. 3). Regardless, we quantify the emission spectra in terms of an effective temperature for ease of comparison. The visible part of the emission is best represented by black-body emission with temperatures above 1200 K. Whenever the excitation is resonant to a vibrational mode, the emission temperature increases by ~200 K (cf. FIG 5b). Indicating, that heating of the molecular backbone also results in heating of the electronic system. The overall very high emission temperature is in stark contrast to the lattice temperature of ~400 K as obtained by upconversion thermometry (See Supplemental Material and Figure S6), rendering thermal emission due to joule heating of the whole molecular entity as the source of the observed supercontinuum questionable to state the least. Altogether, our observations support the identification of the solitary electron system as the source of the supercontinuum.

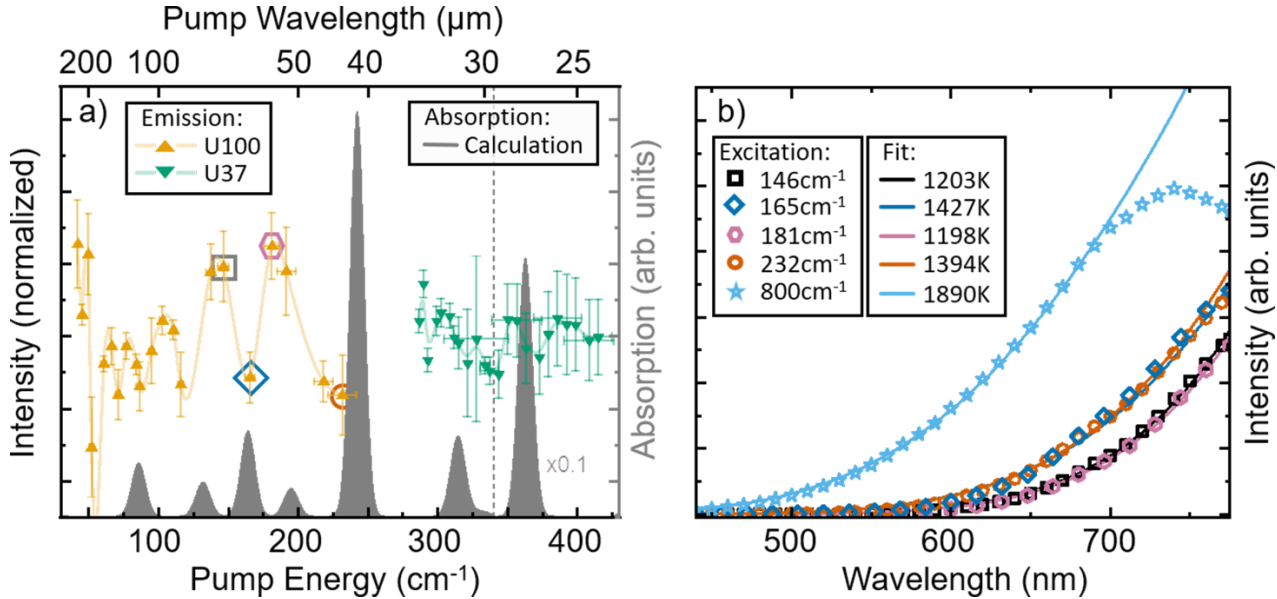

FIG. 5. Pump energy dependent near-infrared emission intensity, and calculated absorption of [(PhSn)$_4$S$_6$]. a) Integrated intensity of the emission measured for excitation using the U37 undulator (green downward triangles) and the U100-undulator (orange upward triangles), respectively. For better comparison both data sets are normalized to the same average value. Infrared absorbance (grey area) of [(PhSn)$_4$S$_6$] computed with DFT methods. For energies above 340 cm$^{-1}$ (dashed line), the calculated spectrum is scaled down by a factor of 10. b) Emission spectra at selected pump energies (open symbols) with fits according to Planck's law representing black-body radiation at the given temperatures (lines).

The calculations further confirm that vibrations above 400 $cm^{-1}$ are predominantly associated with motion of the substituents. These are phenyl rings hosting the delocalized π-electron systems. Consequently, and according to the Born-Oppenheimer approximation, such motion in the ligand system would affect the potential energy surface of the cluster and in turn the distribution of the delocalized π-electron systems. Such a change to the emitting electron distribution could easily result in a change of the spectral distribution of the emission. Thus, we assign the observed change in emission for excitation above 400 $cm^{-1}$ (cf. Fig 3) to induced motion in the delocalized π-electron systems. This is corroborated by a further increase in apparent emission temperature to ~1900 K when exciting at, e.g., 800 $cm^{-1}$ (see Fig 5b). This motion is not present when the excitation energy is not sufficient (<400 $cm^{-1}$) to introduce motion of the ligand system. At the same time, the decoupling of response time scales of the electron system in the ligands and the vibrational motion of the backbone is consistent with the observation of white-light emission rather than heating – and ultimate transformation - of the amorphous cluster compounds.

## IV. SUMMARY

To conclude, we show that advanced nonlinear optical media consisting of amorphous powder of adamantane-like tin-sulfide clusters emit a supercontinuum ranging into the visible spectrum for excitation wavelength in the mid-infrared 8 – 50 µm (1250 $cm^{-1}$ – 200 $cm^{-1}$) and far-infrared 50 – 240 µm (200 $cm^{-1}$ – 42 $cm^{-1}$). Intensity fluctuations of the emission for excitation wavelength above 25 µm (400 $cm^{-1}$) are explained by competing absorption channels. Energy is dissipated into vibrations rather than driving the supercontinuum generation whenever the driving light is resonant to vibrations of the cluster core. DFT-calculation clearly identify the absorption into these modes as the plausible origin of the spectral efficiency dips. The spectral shape of the supercontinuum itself, however, shows only a minor dependence on the driving wavelength as the emission maximum undergoes a red-shift for longer excitation wavelength. Notably, this shift is on the same order-of-magnitude as the corresponding shift of the excitation photon energies. The emission spectrum also contains a significant near and mid infrared contribution.

Overall, our experimental observations corroborate the tentatively ascribed model of charge-acceleration in the delocalized ligand system as motion within the cluster core provide an alternate absorption channel. Therefore, tailored far-infrared or even GHz fields should be capable of generating broadband emission from such amorphous nonlinear media enabling visible white-light generation based on direct electronic driving sources.

## ACKNOWLEDGMENTS

This Work is supported by the German Science Foundation through FOR2824; SC also acknowledges the Heisenberg Programme under contract CH660/08. N.W.R. gratefully acknowledges funding from the Alexander von Humboldt Foundation within the Feodor-Lynen Fellowship program. N.W.R. thanks Per Persson and Milda Pucetaite for access to FTIR spectrometer. Special thanks are owed to the staff at the free electron laser (FELBE) at the Helmholtz-Zentrum-Dresden-Rossendorf (HZDR) for generously supporting the measurements. CA acknowledges funding through the MIUR PRIN (Grant No. 2020JZ5N9M). Calculations for this research were conducted on the Lichtenberg high performance computer of the TU Darmstadt and at the Höchstleistungrechenzentrum Stuttgart (HLRS). The authors furthermore acknowledge the computational resources provided by the HPC Core Facility and the HRZ of the Justus-Liebig-Universität Gießen.